\documentclass[aps,prl, twocolumn, superscriptaddress, longbibliography]{revtex4-1}
\usepackage{amsmath}
\usepackage{amssymb}
\usepackage{graphicx}
\usepackage[caption=false, position=top, singlelinecheck=off, justification=raggedright]{subfig}
\usepackage{color}
\usepackage{pst-node}
\usepackage[unicode]{hyperref}
\hypersetup{unicode=true, colorlinks=true, linkcolor=blue, citecolor=blue}

\begin{document}

\title{Realization of a periodically driven open three-level Dicke model}

\author{Phatthamon Kongkhambut}
\affiliation{Zentrum f\"ur Optische Quantentechnologien and Institut f\"ur Laser-Physik, Universit\"at Hamburg, 22761 Hamburg, Germany}

\author{Hans Ke{\ss}ler}
\affiliation{Zentrum f\"ur Optische Quantentechnologien and Institut f\"ur Laser-Physik, Universit\"at Hamburg, 22761 Hamburg, Germany}

\author{Jim Skulte}
\affiliation{Zentrum f\"ur Optische Quantentechnologien and Institut f\"ur Laser-Physik, Universit\"at Hamburg, 22761 Hamburg, Germany}
\affiliation{The Hamburg Center for Ultrafast Imaging, Luruper Chaussee 149, 22761 Hamburg, Germany}

\author{Ludwig Mathey}
\affiliation{Zentrum f\"ur Optische Quantentechnologien and Institut f\"ur Laser-Physik, Universit\"at Hamburg, 22761 Hamburg, Germany}
\affiliation{The Hamburg Center for Ultrafast Imaging, Luruper Chaussee 149, 22761 Hamburg, Germany}

\author{Jayson G. Cosme}
\affiliation{National Institute of Physics, University of the Philippines, Diliman, Quezon City 1101, Philippines}

\author{Andreas Hemmerich}
\affiliation{Zentrum f\"ur Optische Quantentechnologien and Institut f\"ur Laser-Physik, Universit\"at Hamburg, 22761 Hamburg, Germany}
\affiliation{The Hamburg Center for Ultrafast Imaging, Luruper Chaussee 149, 22761 Hamburg, Germany}

\date{\today}

\begin{abstract}     
A periodically driven open three-level Dicke model is realized by resonantly shaking the pump field in an atom-cavity system. As an unambiguous signature, we demonstrate the emergence of a dynamical phase, in which the atoms periodically localize between the antinodes of the pump lattice, associated with an oscillating net momentum along the pump axis. We observe this dynamical phase through the periodic switching of the relative phase between the pump and cavity fields at a small fraction of the driving frequency, suggesting that it exhibits a time crystalline character.
\end{abstract}


\maketitle
Rapid technological advances have elevated cold-atom systems to preeminent platforms for realizing model systems of quantum-many body dynamics \cite{Bloch2008_2, Bloch2008, Bakr2009, Bloch2012, Gross2017, Bayha2020}. An intriguing sub-class are hybrid light-matter systems, which are composed of cold atoms coupled to an optical cavity, and display a strongly enhanced light-matter interaction, giving access to the physics of strong light-matter coupling and long-range correlations \cite{Ritsch2013, Mivehvar2021}. A specific feature of these platforms is the well controlled dissipation, which allows for fast non-destructive in-situ monitoring of the system dynamics \cite{Black2003, Baumann2011, Klinder2015, Kessler2016, Klinder2016, Georges2017, Kessler2021, Mivehvar2021}. One of the fundamental models for light-matter interaction is the Dicke model \cite{Hepp1973, Kirton2019}. It describes a collection of $N$ two-level systems coupled to a single light mode and displays a phase transition between a normal and a superradiant phase \cite{Hepp1973}. An open version of the Dicke model with a weak dissipation channel is approximately realized by a Bose-Einstein condensate (BEC) placed in a linear standing wave optical cavity and pumped by an optical standing wave oriented perpendicularly with respect to the cavity axis \cite{Nagy2008, Baumann2010, Baumann2011, Klinder2015, Baden2014, Klinder2015, Piazza2015, Lode2017, Kessler2019, Molignini18, Kessler2020, Georges2021, Kessler2021}. The normal phase is characterized by a BEC, light-shifted by the pump potential, with a homogeneous density distribution along the cavity axis and a small number of photons in the cavity that do not display coherence. The superradiant phase shows a density grating enabling pronounced scattering of photons from the pump into the cavity and vice versa. Various extensions of the standard two-level Dicke model have been proposed and realized using atom-cavity systems, such as the spin-1 Dicke model \cite{Zhiqiang2017, Masson2017} and the two-component Dicke model \cite{Chiacchio2019, Buca2019, Dogra2019}, all sharing the coupling of two-level systems to the same monochromatic light mode. 

\begin{figure}[ht!]
\centering
\includegraphics[width=1\columnwidth]{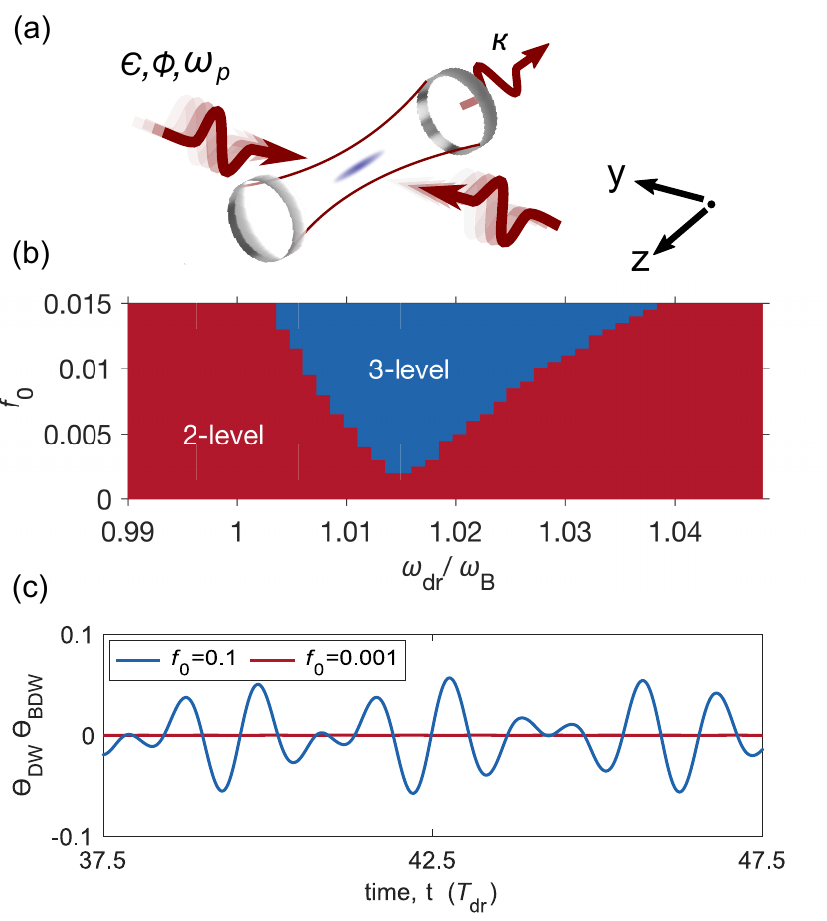}
\caption{(a) Schematic of the transversely pumped and shaken atom-cavity system. A sample of cold atoms is placed in a cavity oriented along the $z$-axis. A standing wave potential is periodically shifted along the $y$-axis using phase modulation techniques. (b) Dynamical phase diagram with two distinct regions: In the red area, the dynamics of the system is captured by a two-level Dicke model. In the blue area, a three-level Dicke model is required. (c) Dynamics of the product of the relevant order parameters for strong driving in the three-level Dicke regime (blue) and for weak driving in the two-level Dicke regime (red). The modulation frequency is $\omega_\mathrm{dr}/2\pi = 9.4~\mathrm{kHz}$ and $\omega_\mathrm{B}/2\pi = 8~\mathrm{kHz}$.}
\label{fig:1} 
\end{figure}

The extension of the Dicke model to the case of three-level systems has been theoretically considered in Refs.\cite{Sung1979, Crubellier1985, Crubellier1986}. A specific example in a ring cavity has been used to experimentally demonstrate subradiance \cite{Wolf2018}. In the present work, we experimentally realize the periodically driven open three-level Dicke model by shaking the standing wave pump potential in an atom-cavity system as depicted in Fig.~\ref{fig:1}(a). It has been predicted in Ref.~\cite{Cosme2019} that this enables a dynamical phase, characterized by atoms periodically localizing between the antinodes of the pump lattice, i.e., on the inter-site bonds, which has been called dynamical bond density wave (DBDW) phase. This DBDW phase exhibits time crystalline character and is a characteristic signature of the periodically driven open three-level Dicke model. Its experimental observation is the central topic of this work. 

We define the three-level Dicke model to describe the interaction between a single quantized light mode and $N$ three-level atoms comprising of energy eigenstates $|1\rangle$, $|2\rangle$, and $|3\rangle$ in a V configuration. Its Hamiltonian is 
\begin{align}\label{eq:ham}
H/ \hbar &= \omega \hat{a}^\dagger \hat{a} + \omega_{\mathrm{12}} \hat{J}^{12}_z   +\omega_{\mathrm{13}}\hat{J}^{\mathrm{13}}_z  \\ \nonumber
&+\frac{2}{\sqrt{N}}\left(\hat{a}^\dagger+\hat{a} \right) \left(\lambda_{12}\hat{J}^{\mathrm{12}}_x + \lambda_{13} \hat{J}^{\mathrm{13}}_x \right).
\end{align}
The bosonic operator $\hat{a}$ ($\hat{a}^\dagger$) annihilates (creates) a photon with frequency $\omega$. The frequency detuning between the lowest energy state $|1\rangle$ and the other two states $|2\rangle$ and $|3\rangle$ are $\omega_{12}$ and $\omega_{13}$, respectively. For small detuning $\omega_{23}$ between the states $|2\rangle$ and $|3\rangle$, i.e., when $\omega_{23}\ll \omega_{12},\omega_{13}$, the only relevant light-matter interactions are those that couple state $|1\rangle$ with states $|2\rangle$ and $|3\rangle$, the strengths of which are given by $\lambda_{12}$ and $\lambda_{13}$, respectively. We introduce the pseudospin operators $\hat{J}^\ell_\mu$ with $\ell \in \{12,13,23\}$, which are related to the eight generators of the SU(3) group \cite{Skulte2021}. Note that the Gell-Mann matrices, the standard representation of the SU(3) group, can be obtained by an appropriate superposition of  $\hat{J}^\ell_\mu$ \cite{Skulte2021}. Eq.~\eqref{eq:ham} is an extended form of the two-component Dicke model \cite{Chiacchio2019, Buca2019, Dogra2019}. However, the latter obeys the SU(2) algebra, while the pseudospin operators in Eq.~\eqref{eq:ham} fulfill the SU(3) algebra, instead. 

To implement the three-level Dicke model, we consider atoms in their electronic ground state occupying the following three momentum states forming a V-shaped level structure (see Fig.~1 in the supplemental material \cite{supp}). The ground state is the so called BEC state  $|\mathrm{BEC}\rangle$ given by the zero momentum state $|0,0\rangle$ with respect to the $yz$-plane, light-shifted by the pump field by an amount $-\epsilon/2$, where $\epsilon$ denotes the potential depth of the pump wave \cite{Skulte2021}. The first excited state is the superposition $\sum_{\nu,\mu\in\{-1,1\}} |\nu \hbar k,\mu \hbar k\rangle$ of the four momentum modes $|\pm \hbar k,\pm \hbar k\rangle$ associated with the $yz$-plane, light-shifted by the pump field by an amount $-3 \epsilon/4$ (Here, $k$ denotes the wave number of the pump field) \cite{Skulte2021}. In view of its spatially varying density $\propto |\cos(ky)\cos(kz)|^2$, it is denoted as the density wave state $|\mathrm{DW}\rangle$. The light-shift for $|\mathrm{DW}\rangle$ is larger compared to that of $|\mathrm{BEC}\rangle$, since the density distribution of $|\mathrm{DW}\rangle$ is localized in the antinodes of the pump field \cite{Skulte2021}. The two states $|\mathrm{BEC}\rangle$ and $|\mathrm{DW}\rangle$ span the matter sector of the regular two-level Dicke model. If $\epsilon$ exceeds a critical value $\epsilon_{\mathrm{crt}}$, $|\mathrm{BEC}\rangle$ acquires an admixture of $|\mathrm{DW}\rangle$. A Bragg grating is thus imprinted upon the density of the $|\mathrm{BEC}\rangle$ state, which via efficient scattering of pump light builds up a coherent intra-cavity light field. The $|\mathrm{BEC}\rangle$ state, thus dressed by the cavity field, is denoted superradiant phase. In this work, we operate either with $\epsilon <\epsilon_{\mathrm{crt}}$ or with $\epsilon$ only very slightly above $\epsilon_{\mathrm{crt}}$, such that the additional dressing by the cavity field is zero or negligibly small. The second excited state is associated with the momentum state superposition $\sum_{\nu,\mu\in\{-1,1\}} \nu |\nu \hbar k,\mu \hbar k\rangle$. This state exhibits the smallest light-shift $-\epsilon / 4$, because its density distribution $\propto |\sin(ky)\cos(kz)|^2$ matches with the nodes of the pump wave \cite{Skulte2021}. This state is called bond density wave (abbreviated $|\mathrm{BDW}\rangle$) as its density maxima coincide with the bonds between two potential minima of the pump wave. We denote the energy separation between $|\mathrm{DW}\rangle$ and $|\mathrm{BEC}\rangle$ as $\hbar \omega_{\mathrm{D}}$, and that between $|\mathrm{BDW}\rangle$ and $|\mathrm{BEC}\rangle$ as $\hbar \omega_{\mathrm{B}}$, respectively. See the supplemental material for a more detailed description \cite{supp}.
  
\begin{figure}[!t]
\centering
\includegraphics[width=1\columnwidth]{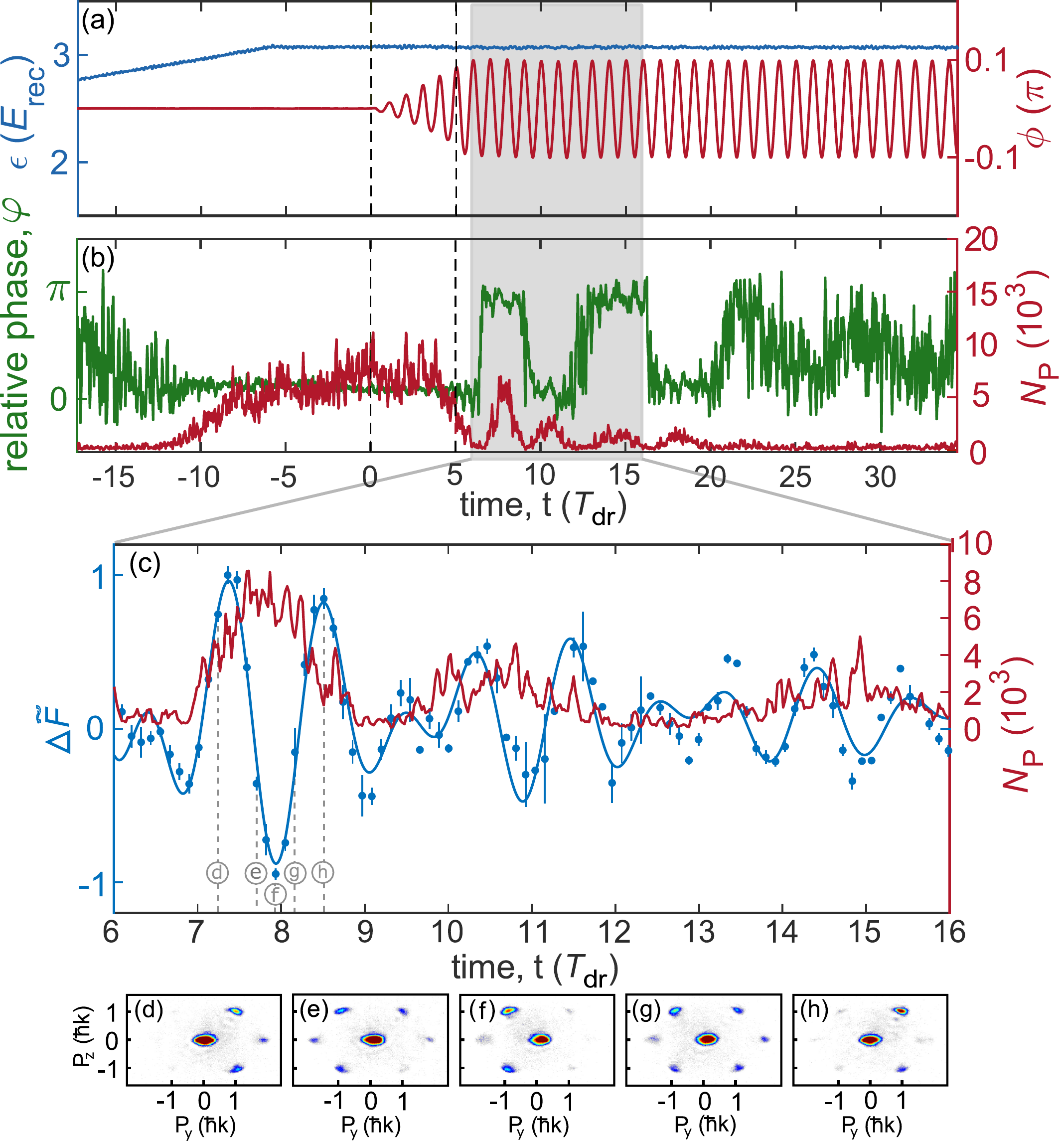}
\caption{{Single-shot realization of DBDW order.}  {(a)} Time sequence for the pump lattice depth (blue) and the phase $\phi$ of the pump field (red) with modulation strength $f_\mathrm{0}=$0.1$\,\pi$ and a modulation frequency $\omega_\mathrm{dr}=11.5\,$kHz.  {(b)} Phase difference $\varphi$ between the pump and intra-cavity field (green trace) and photon number $N_\mathrm{P}$ in the cavity (red trace). The dashed vertical lines mark the time interval during which the modulation strength is increased. The gray shaded area shows the time window for the zoom presented in (c).  {(c)} The red trace repeats the intracavity photon number $N_\mathrm{P}$ from (b). The blue data points plot the product $\Theta_{\mathrm{DW}} \times \Theta_{\mathrm{BDW}} $, approximately given by $\Delta \tilde{F}$ (see also Fig.~\ref{fig:1}(c)). Each data point is averaged over 5 realizations. The solid line shows a fit with a product of two harmonic oscillations. {(d)-(h)} Single-shot momentum distributions recorded at the times marked in (c).}
\label{fig:2} 
\end{figure}

In the atom-cavity implementation of the standard Dicke model, $|\mathrm{BDW}\rangle$ is not coupled to $|\mathrm{BEC}\rangle$ and hence can be dropped. To implement a coupling between $|\mathrm{BDW}\rangle$ and $|\mathrm{BEC}\rangle$, the transverse pump lattice is periodically shaken in space \cite{Cosme2019}. In Ref.~\cite{Skulte2021}, we show that the Hamiltonian for the shaken atom-cavity system can be mapped onto a parametrically driven version of the three-level Dicke model.
\begin{align}\label{eq:ham2}
H/ \hbar &= \omega \hat{a}^\dagger \hat{a} +\hat{J}^{\mathrm{D}}_z  \omega_{\mathrm{D}} +\hat{J}^{\mathrm{B}}_z  \omega_{\mathrm{B}}
 +2\phi(t) \left( \omega_\mathrm{B}-\omega_\mathrm{D} \right) \hat{J}^{\mathrm{DB}}_x \nonumber \\ 
&+\frac{2\lambda}{\sqrt{N}}\left(\hat{a}^\dagger+\hat{a} \right) \left(\hat{J}^{\mathrm{D}}_x -\phi(t) \hat{J}^{\mathrm{B}}_x \right), 
\end{align}
where $\phi(t)=f_0 \sin(\omega_\mathrm{dr} t)$ is the time-dependent spatial phase of the pump lattice introduced by the shaking protocol, and $\lambda$ is the overall coupling strength parameter. The pseudospin operators $\hat{J}^{\mathrm{D}}_\mu$ and  $\hat{J}^{\mathrm{B}}_\mu$ with $\mu \in\{x,y,z\}$ are directly associated with the $|\mathrm{DW}\rangle$ and the $|\mathrm{BDW}\rangle$ states via the relations to their order parameters $\Theta_{\mathrm{DW}} \equiv \langle \cos(ky)\cos(kz) \rangle = \langle \hat{J}^{\mathrm{D}}_x \rangle$ and $\Theta_{\mathrm{BDW}} \equiv \langle \sin(ky)\cos(kz) \rangle = \langle \hat{J}^{\mathrm{B}}_x \rangle$, respectively. Comparing Eqs.~\eqref{eq:ham} and \eqref{eq:ham2}, we identify $\hat{J}^{12}_\mu=\hat{J}^\mathrm{D}_\mu  $, $\hat{J}^{13}_\mu =\hat{J}^\mathrm{B}_\mu$, $ \hat{J}^{23}_{\mu}=\hat{J}^{\mathrm{DB}}_{\mu}$, $\omega_{12} = \omega_{\mathrm{D}}$,  $\omega_{13} = \omega_{\mathrm{B}}$, $\lambda_{12}=\lambda$, and a time-dependent light-matter coupling $\lambda_{13} = -\phi(t)\lambda$. Moreover, in Eq.~\eqref{eq:ham2}, the standing wave potential of the pump introduces an additional albeit negligible term proportional to $\hat{J}^{\mathrm{DB}}_x$, which couples $|\mathrm{DW}\rangle$ and $|\mathrm{BDW}\rangle$ \cite{Skulte2021}.

\begin{figure}[!t]
\centering
\includegraphics[width=0.9\columnwidth]{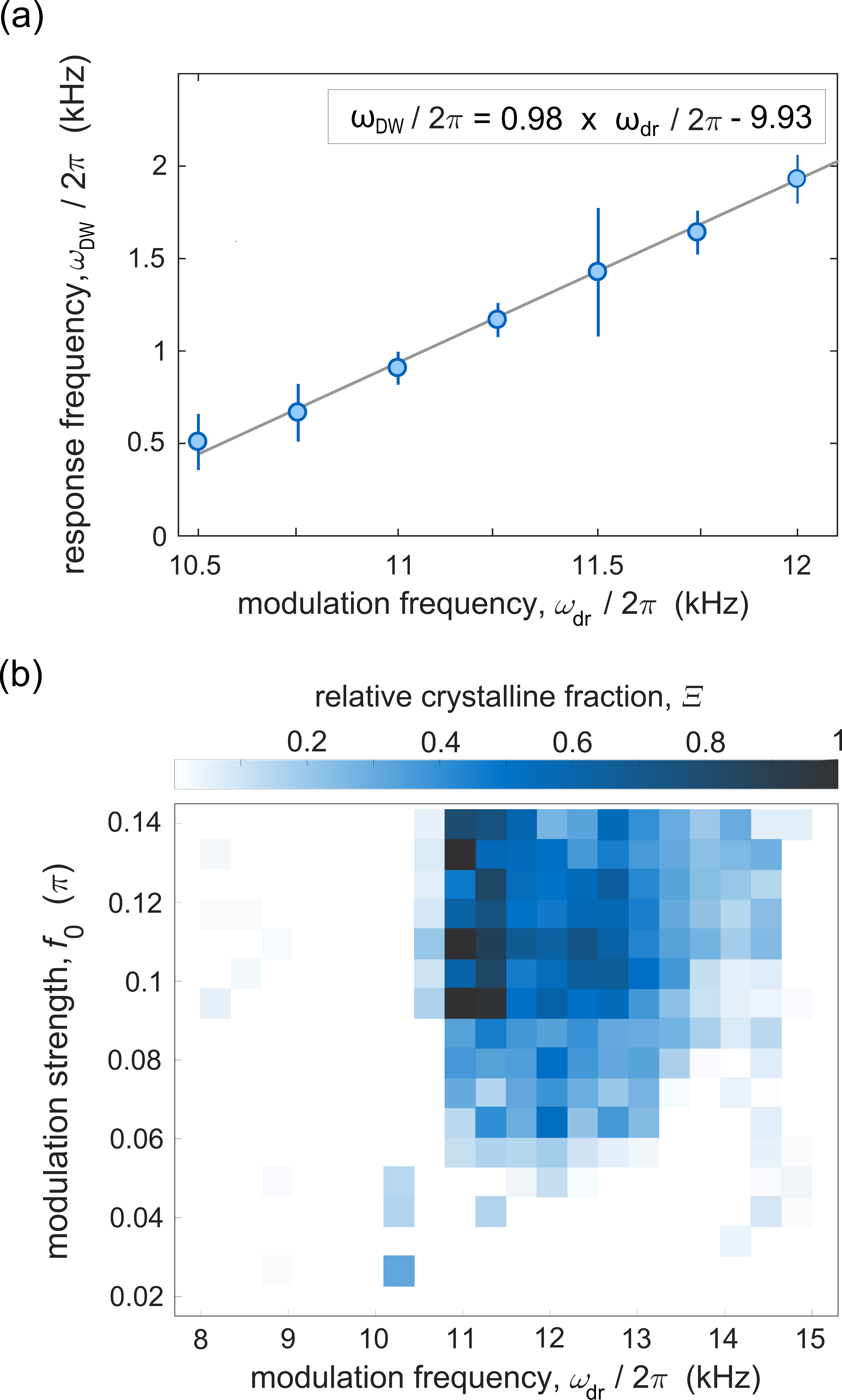}
\caption{(a) $\omega_\mathrm{DW}$ is plotted against $\omega_\mathrm{dr}$. $\omega_\mathrm{DW}$ is extracted by the position of a Gaussian fit of the amplitude spectrum calculated from the measured time evolution of the phase difference between the pump and cavity fields $\varphi$. Each data point is averaged over hundreds of realizations with different modulation strength $f_\mathrm{0}$ and fixed $\omega_\mathrm{dr}$. The gray line is a linear fit yielding the result shown in the plot legend. (b) The relative crystalline fraction $\Xi$ is plotted as a function of the modulation frequency $\omega_\mathrm{dr}$ and strength $f_\mathrm{0}$. The diagram is constructed by dividing the parameter space into $20\times16$ plaquettes and averaging over multiple experimental runs within each.}
\label{fig:3} 
\end{figure} 

For driving frequencies $\omega_{\mathrm{dr}}$ slightly above $\omega_{\mathrm{B}}$, the DBDW phase shows periodic oscillations of $\Theta_{\mathrm{BDW}}$ and $\Theta_{\mathrm{DW}}$ around zero with frequencies $\omega_{\mathrm{BDW}}$ and $\omega_{\mathrm{DW}}$, respectively. Theory predicts the relation $\omega_\mathrm{DW} = \omega_\mathrm{dr} - \omega_\mathrm{BDW}$ such that $\omega_\mathrm{DW}$ is not an integer fraction of the driving frequency $\omega_\mathrm{dr}$ \cite{Cosme2019}. This is a hallmark of an incommensurate time crystal \cite{Cosme2019}. Thus, the long-time average of $\Theta_{\mathrm{DW}}$ is zero in the three-level Dicke region of the dynamical phase diagram, while it is nonzero in the two-level Dicke region for an initial superradiant phase. This behavior is captured in Fig.~\ref{fig:1}(b), which shows the time-averaged value of $\langle \hat{J}^{\mathrm{D}}_x \rangle/N \equiv  j^{\mathrm{D}}_x $ obtained by solving the equations of motion corresponding to Eq.~\eqref{eq:ham2} in the semiclassical limit of a large atom number \cite{Skulte2021}.

The DBDW dynamics may be experimentally studied via the product of the order parameters $\Theta_{\mathrm{DW}} \times \Theta_{\mathrm{BDW}}$, which can be approximately measured by the normalized occupation imbalance $\Delta \tilde{F} \equiv$ $(F_\mathrm{+1,\pm1}\, - \,F_\mathrm{-1,\pm1}) / (F_\mathrm{+1,\pm1}\, - \,F_\mathrm{-1,\pm1})_\mathrm{max}$, where $F_\mathrm{\pm1,\pm1}$ denotes the population of the momentum state $|\pm \hbar k,\pm \hbar k\rangle$ (see the supplemental material for details \cite{supp}). In the standard Dicke model realized for off-resonant driving, $\Theta_{\mathrm{BDW}}\approx 0 $ and $\Delta \tilde{F}$  is negligible. On the other hand, for driving frequencies $\omega_{\mathrm{dr}}$ slightly above $\omega_{\mathrm{B}}$, a beating signal is expected in $\Theta_{\mathrm{DW}} \times \Theta_{\mathrm{BDW}}$ (see Fig.~\ref{fig:1}(c)), which can be observed via $\Delta \tilde{F}$. Furthermore, the periodic switching of $\Theta_{\mathrm{DW}}$ in the three-level model amounts to a periodic switching of the experimentally observable relative phase of the pump and the cavity fields $\varphi \equiv \mathrm{arg}(\langle \hat{a} \rangle)$ between $0$ and $\pi$. 

In our experiment, a BEC of $^{87}$Rb atoms is superimposed with the fundamental mode of a high-finesse optical cavity pumped by a retro-reflected laser beam at wavelength $\lambda_\mathrm{P}=803\,$nm. The resulting optical pump lattice has a depth $\epsilon$ and is aligned perpendicular to the cavity axis, as depicted in Fig.~\ref{fig:1}(a). The cavity has a field decay rate $\kappa=2\pi\times3.6\,$kHz comparable to the recoil frequency $\omega_{\mathrm{rec}}\equiv \hbar k^2/2m$ ($m =$ atomic mass), such that the cavity field and the atomic density distribution evolve on similar timescales. This leads to a retarded infinite-range cavity-mediated interaction between the atoms \cite{Klinder2016}. The system realizes the Dicke phase transition from a homogeneous BEC to a superradiant phase if $\epsilon$ exceeds a critical strength. The $\mathbb{Z}_2$ symmetry is spontaneously broken, when the atoms localize at either the even or odd sites of a two dimensional chequerboard optical lattice formed by the interference between the pump and intra-cavity fields. The two symmetry broken states can be distinguished by the relative phase difference $\varphi$ between the pump and intra-cavity light fields using a balanced heterodyne detection of the cavity field. The appearance of the superradiant phase can be detected in-situ by the observation of a nonzero cavity mode occupation $N_\mathrm{P}$ (see red line in Fig.~\ref{fig:2}(b)), the locking of the relative $\varphi$ to zero or $\pi$ (see green line Fig.~\ref{fig:2}(b)), or in a destructive way through a nonzero occupation of the $\{p_y,p_z\}=\{\pm\hbar k,\pm\hbar k\}$ modes in a momentum spectrum (see Fig.~\ref{fig:2}(g)).

The experimental sequence proceeds as follows: We prepare the system in the BEC phase or in the superradiant phase close to the phase boundary towards the BEC phase, followed by a 500\,$\mu$s long waiting period to let the system reach its steady state. Then, we shake the pump potential by modulating the phase of the pump field using an electro-optical modulator. The modulation strength $f_\mathrm{0}$ is linearly increased to its desired value within 500\,$\mu$s and kept constant for 6.5~ms. A typical sequence of the pump protocol is presented in Fig.~\ref{fig:2}(a). Resonant driving induces a switching of the system between the two possible sublattices of the superradiant phase at a frequency $\omega_\mathrm{DW}$ and the intra-cavity photon number pulsates at a rate of $2\,\omega_\mathrm{DW}$. This behavior is exemplified in the green and red curves in Fig.~\ref{fig:2}(b).

In Fig.~\ref{fig:3}(a), we plot $\omega_\mathrm{DW}$ as a function of $\omega_\mathrm{dr}$ and average each data point over 100 experimental runs including different modulation strength $f_\mathrm{0}$. The solid gray trace shows a linear fit. We find good agreement with the theoretical prediction $\omega_\mathrm{DW} = \omega_\mathrm{dr} - \omega_\mathrm{BDW}$ of Ref.~\cite{Cosme2019}. In the supplemental material, we present a similar plot for fixed $\omega_\mathrm{dr}$ and varying $f_\mathrm{0}$ to show that the dependence of $\omega_\mathrm{DW}$ on $f_\mathrm{0}$ is very weak and negligible within the experimental precision \cite{supp}. From the linear fit in Fig.~\ref{fig:3}(a), we extract the value of the parametric resonance as $\omega_\mathrm{BDW} = 9.93\pm0.30\,$kHz. In the supplemental material, we also present an alternative protocol for measuring $\omega_\mathrm{BDW}$ from the depletion of the cavity field for resonant modulation \cite{supp}. In Fig.~\ref{fig:3}(b), we present the dynamical phase diagram, highlighting the DBDW order obtained from measuring the relative crystalline fraction $\Xi$ quantified by the color scale. The relative crystalline fraction is a quantity commonly used in studies of time crystals. Here, we define it as the amplitude of the Fourier spectrum, calculated from the relative phase $\varphi$, at the expected DW frequency $\omega_\mathrm{DW}$, rescaled by its maximum value across the parameter space spanned in the phase diagram \cite{Kessler2021}. The observed DW frequency follows the linear equation $\omega_\mathrm{DW} =\xi \times\omega_\mathrm{dr} - \omega_\mathrm{BDW}$ with $\xi$ determined according to the linear fit in Fig.~\ref{fig:3}(a) as 0.98, i.e., very close to the expected value of unity. This incommensurate subharmonic response of the system with respect to the modulation frequency $\omega_\mathrm{dr}$ is observed within a broad area of the dynamical phase diagram in Fig.~\ref{fig:3}(b). In the supplemental material, we present the robustness of the subharmonic response against temporal noise, which corroborates the classification of this dynamical phase as an incommensurate time crystal.

\begin{figure}[!t]
\centering
\includegraphics[scale=0.35]{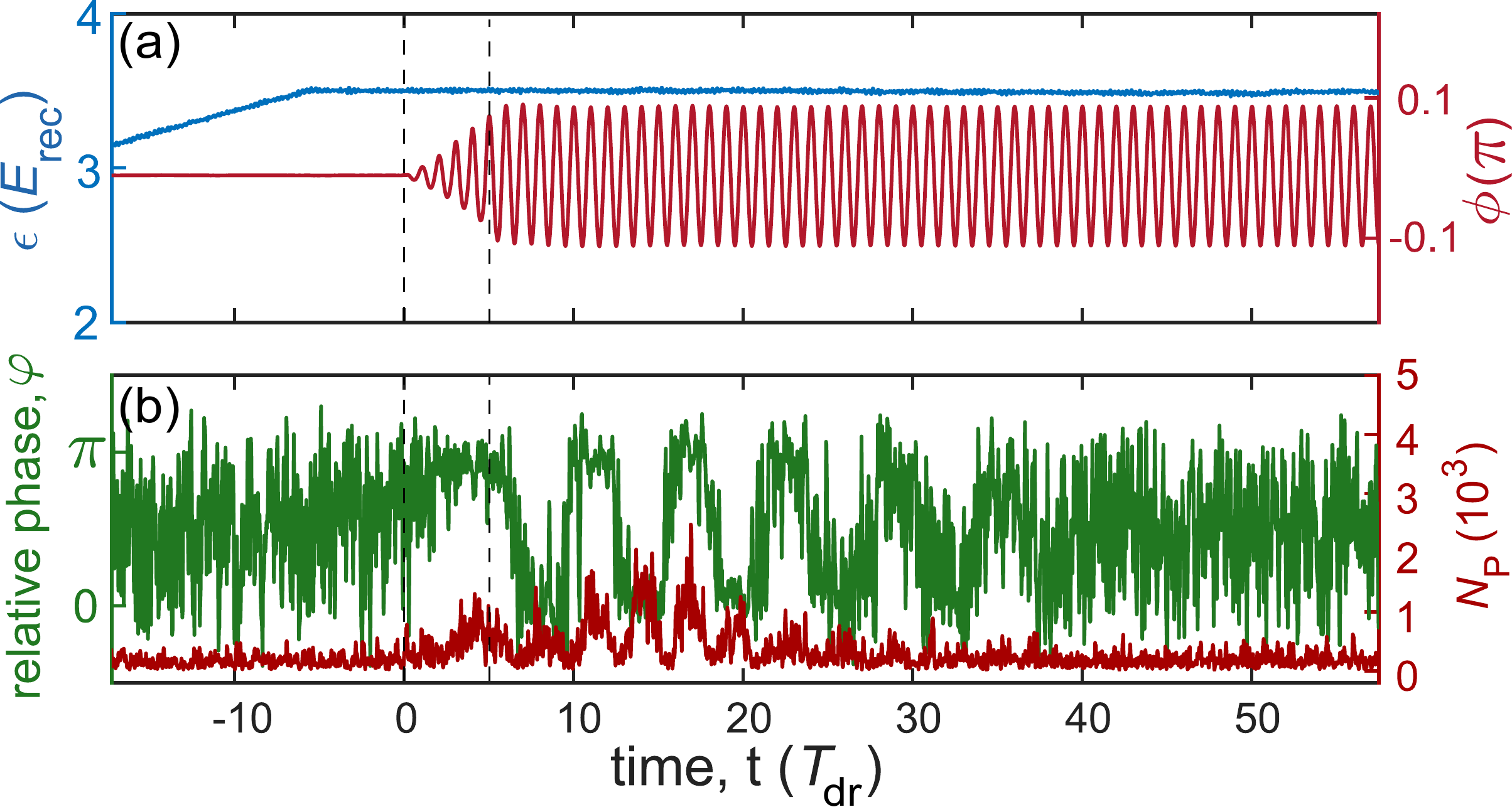}
\caption{{Dynamics in the three-level Dicke regime using an initial homogeneous BEC state.} {(a)} Time sequence for the pump lattice depth (blue) and the phase $\phi$ of the pump field (red) with modulation strength $f_\mathrm{0} = 0.1\,\pi$ and a modulation frequency $\omega_\mathrm{dr}=11.5\,$kHz.  {(b)} The phase difference $\varphi$ between the pump and intracavity field is plotted in green and the photon number $N_\mathrm{P}$ in the cavity in red.}
\label{fig:4} 
\end{figure}

Finally, we discuss the observed dynamics of the momentum imbalance parameter $\Delta \tilde{F}$ related to the calculations in Fig.~\ref{fig:1}(c). The oscillation frequencies $\omega_\mathrm{DW}$ and $\omega_\mathrm{BDW}$ are extracted from the data in Fig.~\ref{fig:2}(c) using $f(t)=\exp(-\tau t)\,A\sin(\omega_\mathrm{BDW}t+\alpha) \,\sin(\omega_\mathrm{DW}t)$ as a fit function. Here, $\tau$ is the decay rate of $N_\mathrm{P}$ and $A$ is an overall amplitude parameter. This measurement demonstrates a third option for measuring $\omega_\mathrm{BDW}$. However, since recording the momentum spectra is a destructive measurement, this method is much more time consuming than simply detecting the light leaking out of the cavity, which makes it extremely difficult to explore large areas in the parameter space. Nevertheless, we repeated this measurement for a second set of modulation parameters shown in the supplemental material \cite{supp}. The frequency $\omega_\mathrm{BDW}$ is independent of $\omega_\mathrm{dr}$ and we measure $\omega_\mathrm{BDW} = 2\pi \times 9.8 \pm0.1\,$kHz. For a driving frequency of  $\omega_\mathrm{dr}=11.5\,$kHz, we measure a slow oscillation frequency of $\omega_\mathrm{DW}=2\pi \times 1.8\pm0.1\,$kHz (see Fig.~\ref{fig:2}(c)), which agrees well with the theoretical prediction of $\omega_\mathrm{DW}=\omega_\mathrm{dr}-\omega_\mathrm{BDW}=2\pi\times(11.5 - 9.8)\,\mbox{kHz}=2\pi\times1.7\,$kHz.

While we have mostly focused on the case when initially the superradiant state is prepared, we have also confirmed that it is possible to enter the three-level regime heralded by the emergence of the DBDW phase by initializing with the homogeneous BEC or normal phase as exemplified in Fig.~\ref{fig:4}.

The finite lifetime of the emergent DBDW phase in our experiment can be mainly attributed to atom losses. Furthermore, we note that our numerical simulations indicate that contact interactions \cite{Cosme2019} and larger detunings $\omega_\mathrm{dr} - \omega_\mathrm{B}$ \cite{supp} decrease the lifetime of the time crystalline response. In the experiment, however, it is difficult to quantitatively separate the effects of atom losses, contact interaction, and detuning from the resonance.

In conclusion, we have realized a periodically driven open three-level Dicke model using a resonantly shaken atom-cavity system. As the main signature of the three-level Dicke model, we have demonstrated the emergence of a dynamical bond density wave phase. When prepared in the three-level Dicke regime, our system realizes an incommensurate time crystal, whereby the atoms periodically self-organize along the bonds of the pump lattice. This advances the understanding of cavity-BEC systems beyond the standard two-level Dicke model, and broadens the scope of dynamically induced many-body states in this and related hybrid light-matter systems. 

\begin{acknowledgments}
We thank G. Homann and L. Broers for useful discussions.
This work is funded by the Deutsche Forschungsgemeinschaft (DFG, German Research Foundation) – SFB-925 – project 170620586 and the Cluster of Excellence “Advanced Imaging of Matter” (EXC 2056), Project No. 390715994. J.S. acknowledges support from the German Academic Scholarship Foundation.
\end{acknowledgments}


\bibliography{references_BDW}

\end{document}


\title{Supplemental Material for\\Realization of a periodically driven open three-level Dicke model}

\author{Phatthamon Kongkhambut}
\affiliation{Zentrum f\"ur Optische Quantentechnologien and Institut f\"ur Laser-Physik, Universit\"at Hamburg, 22761 Hamburg, Germany}

\author{Hans Ke{\ss}ler}
\affiliation{Zentrum f\"ur Optische Quantentechnologien and Institut f\"ur Laser-Physik, Universit\"at Hamburg, 22761 Hamburg, Germany}

\author{Jim Skulte}
\affiliation{Zentrum f\"ur Optische Quantentechnologien and Institut f\"ur Laser-Physik, Universit\"at Hamburg, 22761 Hamburg, Germany}
\affiliation{The Hamburg Center for Ultrafast Imaging, Luruper Chaussee 149, 22761 Hamburg, Germany}

\author{Ludwig Mathey}
\affiliation{Zentrum f\"ur Optische Quantentechnologien and Institut f\"ur Laser-Physik, Universit\"at Hamburg, 22761 Hamburg, Germany}
\affiliation{The Hamburg Center for Ultrafast Imaging, Luruper Chaussee 149, 22761 Hamburg, Germany}

\author{Jayson G. Cosme}
\affiliation{National Institute of Physics, University of the Philippines, Diliman, Quezon City 1101, Philippines}

\author{Andreas Hemmerich}
\affiliation{Zentrum f\"ur Optische Quantentechnologien and Institut f\"ur Laser-Physik, Universit\"at Hamburg, 22761 Hamburg, Germany}
\affiliation{The Hamburg Center for Ultrafast Imaging, Luruper Chaussee 149, 22761 Hamburg, Germany}

\date{\today}

\maketitle

\section{Experimental details}
The experimental set-up, as sketched in Fig.~1(a) in the main text, is comprised of a magnetically trapped BEC of $N_\mathrm{a} = 65\times 10^3$  $^{87}$Rb atoms, dispersively coupled to a narrow-band high-finesse optical cavity. The cavity field has a decay rate of $\kappa=~2\pi \times$3.6~kHz, which almost equals the recoil frequency $\omega_\mathrm{rec} = E_\mathrm{rec}/\hbar  =~2\pi \times$3.55~kHz. The wavelength of the pump laser is $\lambda_\mathrm{P} = 803\,$nm, which is red detuned with respect to the relevant atomic transition of $^{87}$Rb at 795~nm. The maximum light shift per atom is $U_0 = -2\pi \times 0.36~\mathrm{Hz}$. We fix the effective detuning to $\delta_{\mathrm{eff}} \equiv \delta_{\mathrm{C}} - (1/2)N_\mathrm{a} U_0=-2 \pi \times 18.5~\mathrm{kHz}$, where $\delta_{\mathrm{C}} = \omega_{\mathrm{P}}-\omega_{\mathrm{C}}$ is the pump-cavity detuning. A typical experimental sequence starts by preparing the system in the superradiant phase. This is achieved by linearly increasing the pump strength $\epsilon$ from zero to its final value $\epsilon_\mathrm{0}=3.3~E_\mathrm{rec}$ in 10~ms at a fixed effective pump-cavity detuning $\delta_\mathrm{eff}=-2\pi \times$18.5~kHz.

\section{Three-level system}
\begin{figure}[!htbp]
\centering
\includegraphics[width=0.6\columnwidth]{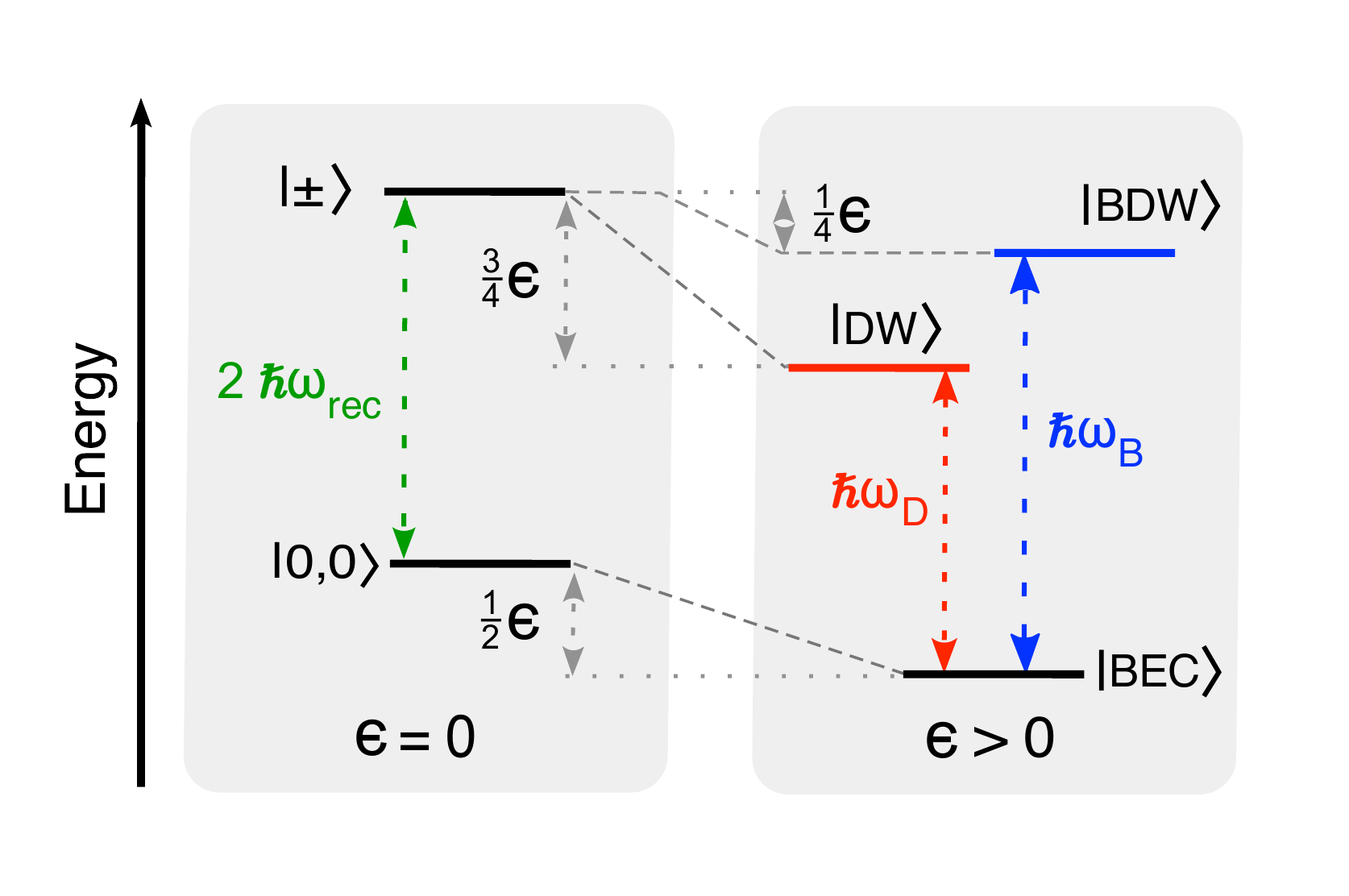}
\caption{{V-shaped three-level system.} On the left, the case of zero pump field strength $\epsilon = 0$ is shown with the zero momentum ground state $|0,0\rangle$ and two degenerate excited states $|+\rangle \equiv$ $\sum_{\nu,\mu\in\{-1,1\}} |\nu \hbar k,\mu \hbar k\rangle$ and $|-\rangle \equiv$ $\sum_{\nu,\mu\in\{-1,1\}} \nu |\nu \hbar k,\mu \hbar k\rangle$ associated with an energy $2 \hbar \omega_{\mathrm{rec}}$ above that of $|0,0\rangle$ ($\hbar \omega_{\mathrm{rec}} = $ recoil energy). For $\epsilon > 0$, these bare states acquire different light-shifts giving rise to the modified states $|\mathrm{BEC}\rangle$, $|\mathrm{DW}\rangle$, and $|\mathrm{BDW}\rangle$. They span a three-level system with the resonance frequencies $\omega_\mathrm{D}$ and $\omega_\mathrm{B}$ for the left and right leg of the V-shaped coupling scheme, respectively.}
\label{sfig:8} 
\end{figure}
Consider atoms in their electronic ground state. For each atom, a V-shaped three-level system arises as sketched in Fig.~\ref{sfig:8}. For a vanishing pump field $\epsilon = 0$, the ground state is the bare zero momentum state $|0,0\rangle$ and we consider two degenerate excited momentum states given by the even and odd superpositions $|+\rangle \equiv$ $\sum_{\nu,\mu\in\{-1,1\}} |\nu \hbar k,\mu \hbar k\rangle$ and $|-\rangle \equiv$ $\sum_{\nu,\mu\in\{-1,1\}} \nu |\nu \hbar k,\mu \hbar k\rangle$, respectively. Here, $|\pm \hbar k,\pm \hbar k\rangle$ denotes the momentum eigenstates with $\pm \hbar k$ momentum along the pump axis ($y$-axis) and $\pm \hbar k$ momentum along the cavity axis ($z$-axis). As shown in Fig.~\ref{sfig:8}, in the presence of a pump field, these states acquire light-shifts of different sizes, giving rise to the three modified states $|\mathrm{BEC}\rangle$, $|\mathrm{DW}\rangle$ and $|\mathrm{BDW}\rangle$. The $|\mathrm{BEC}\rangle$ state is associated with the zero momentum state $|0,0\rangle$, and hence a homogeneous density distribution. The light-shift for this state is $-\epsilon / 2$ with $\epsilon$ denoting the potential depth of the pump standing wave \cite{Skulte2021}. $|\mathrm{DW}\rangle$ is associated with the bare momentum state $|+\rangle$ and therefore a density distribution proportional to $|\cos(ky)\cos(ky)|^2$. This distribution is localized in the antinodes of the pump wave and thus possesses a larger light-shift $-3 \epsilon / 4$ \cite{Skulte2021}. Finally, $|\mathrm{BDW}\rangle$ is associated with $|-\rangle$ and therefore a density distribution proportional to $|\sin(ky)\cos(ky)|^2$, which matches with the nodes of the pump wave and hence possesses the smallest light-shift $-\epsilon / 4$ \cite{Skulte2021}. 

The preceding discussion strictly applies, if $\epsilon$ remains below a critical value $\epsilon_{\mathrm{crt}}$, beyond which the $|\mathrm{BEC}\rangle$ state undergoes a phase transition to the superradiant state of the regular two-level Dicke model. In particular, above $\epsilon_{\mathrm{crt}}$ a coherent intra-cavity field arises, which mixes $|\mathrm{BEC}\rangle$ and $|\mathrm{DW}\rangle$ and adds additional light-shifts to these states. In the present work, we operate either below or slightly above $\epsilon_{\mathrm{crt}}$, where this additional mixing and the associated light-shifts are assumed sufficiently small to be neglected. 

\section{Atom-Cavity model}
Considering only the pump and cavity directions and neglecting contact interactions between the atoms, the shaken atom-cavity system can be modeled by the many-body Hamiltonian
\begin{align}\label{eq:hamac}
\hat{H}/\hbar = - \delta_{\mathrm{C}} \hat{a}^\dagger \hat{a} + \int dy dz \hat{\Psi}^\dagger (y,z) \biggl[ -\frac{\hbar}{2m} \nabla^2 &-\omega_\mathrm{rec}\epsilon \cos^2(ky+\phi(t)) + U_0  \hat{a}^\dagger \hat{a} \cos^2(kz)  \\ \nonumber
& -\sqrt{\omega_\mathrm{rec}|U_0|\epsilon_p}\cos(ky+\phi(t))\cos(kz)(a^\dagger+a)  \biggr] \hat{\Psi}(y,z),
\end{align}
where $\delta_{\mathrm{C}}$ is the pump-cavity detuning, $U_0<0$ is the maximum light shift per atom, and $\epsilon$ is the pump intensity in units of the recoil energy $E_{\mathrm{rec}}$. This Hamiltonian can be mapped onto the driven open three-level Dicke model (cf. main text) by considering only the five lowest momentum modes of the atoms, $|0,0\rangle$ and $|\pm \hbar k, \pm \hbar k\rangle $, where $k=2\pi/\lambda_\mathrm{P}$ is the wavenumber of the pump (see \cite{Skulte2021} for details). This assumption is valid when the occupations of higher momentum modes are kept negligible, by initializing the system close to the phase boundary between the homogeneous $|\mathrm{BEC}\rangle$ state and the superradiant phase. The matter sector of the superradiant phase, then mainly consists of the $|\mathrm{BEC}\rangle$ state with a small admixture of the $|\mathrm{DW}\rangle$ state, which exhibits a density modulation $\propto |\sin(ky)\cos(kz)|^2$ and hence, the bosonic atomic field operator can be expanded as $ \hat{\Psi}(y,z) \sim \hat{c}_1 + \hat{c}_2 2\cos(ky)\cos(kz)$ \cite{Mivehvar2021}. The Schwinger boson representation can be used to map the transversely pumped atom-cavity Hamiltonian onto the standard two-level Dicke model \cite{Baumann2010,Mivehvar2021}. The order parameter for the $|\mathrm{DW}\rangle$ state is $\Theta_{\mathrm{DW}} = \langle \cos(ky)\cos(kz) \rangle$ in bosonic operator representation, while it is $\Theta_{\mathrm{DW}} = \langle \hat{J}^{\mathrm{D}}_x \rangle$ in pseudospin representation. In our experiment, periodic shaking allows for occupation of the $|\mathrm{BDW}\rangle$ state, which displays a density modulation $ \propto |\sin(ky)\cos(kz)|^2$. The order parameter for $|\mathrm{BDW}\rangle$ is either $\Theta_{\mathrm{BDW}} = \langle \sin(ky)\cos(kz) \rangle$ or $\Theta_{\mathrm{BDW}} = \langle \hat{J}^{\mathrm{B}}_x \rangle$. Taking this into account, the atomic field operator should be extended as $ \hat{\Psi}(y,z) \sim \hat{c}_1 + \hat{c}_2 2\cos(ky)\cos(kz) + \hat{c}_3 2\sin(ky)\cos(kz)$. The driven three-level Dicke model in the main text can then be obtained using an extended Schwinger boson representation that includes this new mode \cite{Skulte2021}. The mapping leads to an effective cavity field frequency of $\omega = (3U_0 N)/4-\delta_\mathrm{C}$. The strength of the light-matter interaction is $\lambda / \sqrt{N}=-\sqrt{\omega_\mathrm{rec}\epsilon_\mathrm{p} |U_0|}/{2}$. Moreover, we obtain $\omega_{\mathrm{D}} =  2 \omega_\mathrm{rec}(1-\epsilon_\mathrm{p}/8)$ and $\omega_{\mathrm{B}} =  2 \omega_\mathrm{rec}(1+\epsilon_\mathrm{p}/8)$ (see Eq.~(2) of the main text). These frequency shifts are depicted in Fig.~\ref{sfig:8}.

\section{Dependence of the density wave frequency on the modulation strength}
\begin{figure}[!htbp]
\centering
\includegraphics[width=0.5\columnwidth]{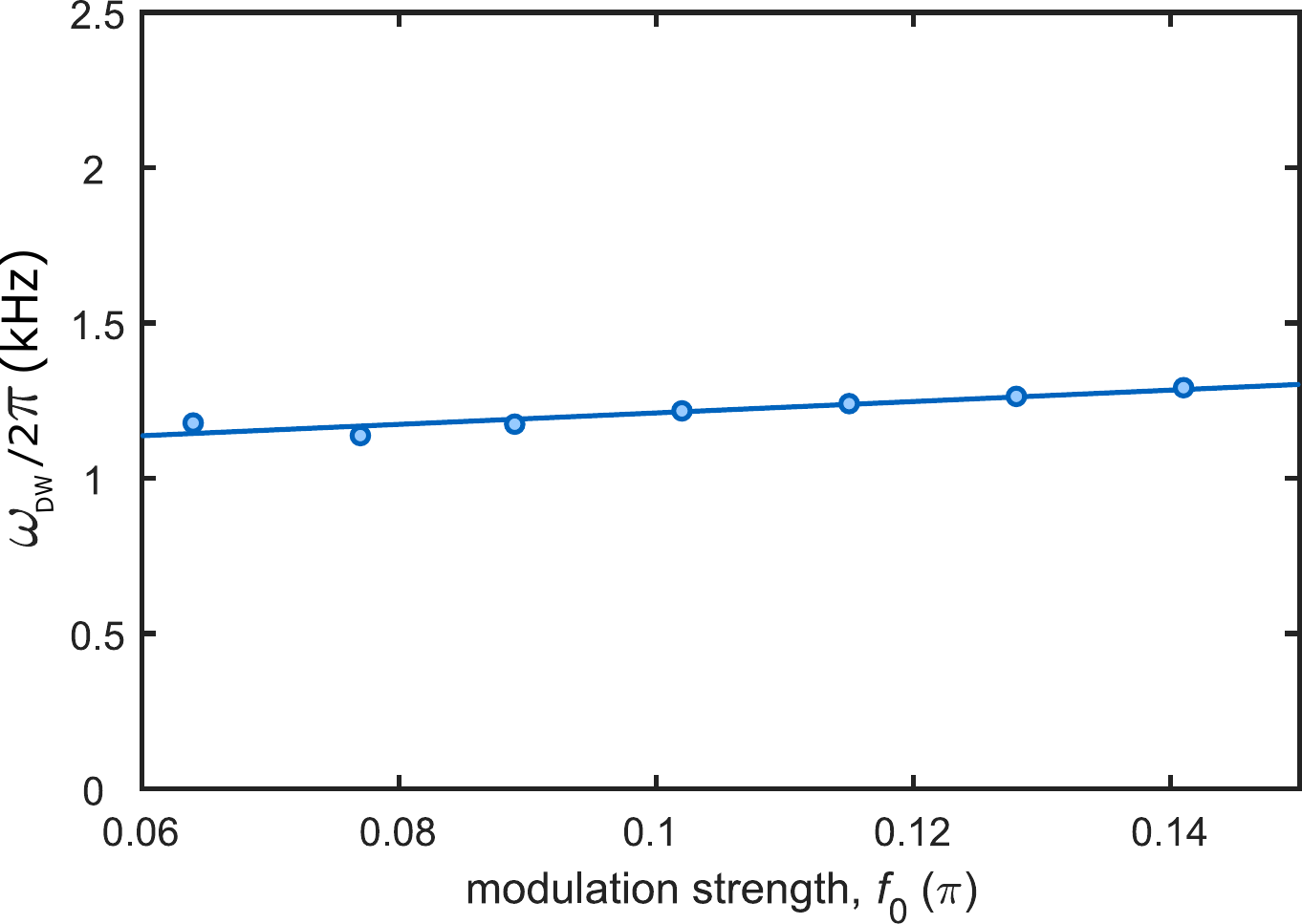}
\caption{{Response frequency $\omega_\mathrm{DW}$ versus modulation strength.} {$\omega_\mathrm{DW}$ is plotted against $f_\mathrm{0}$ for fixed $\omega_\mathrm{dr}$ = $2\pi \times 11.5\,$kHz, using the protocol described in Fig.~3(a) of the main text.}}
\label{sfig:1} 
\end{figure}

As shown in Fig.~\ref{sfig:1}, the density wave frequency $\omega_\mathrm{DW}$ depends weakly on the modulation strength $f_\mathrm{0}$ but the slope is much smaller as for the dependence on the modulation frequency $\omega_\mathrm{dr}$ and we neglect this effect in the construction of the phase diagram in Fig.~3(b) of the main text. It can be explained as follows. Due to the modulation the atoms are sitting on the slope of the light-induced potential and they effectively feel a slightly weaker pump lattice depth $\epsilon$. This effect increases with increasing $f_\mathrm{0}$. Since the position of the bond density wave resonance $\omega_\mathrm{BDW}$ shifts to lower values for smaller $\epsilon$ and $\omega_\mathrm{DW} = \omega_\mathrm{BDW} - \omega_\mathrm{dr}$, the density wave frequency response $\omega_\mathrm{DW}$ increases with increasing modulation strength $f_\mathrm{0}$. 

\section{Measuring the position of the parametric resonance by the depletion of the intracavity field.}
As mentioned in the main text, there are three possibilities to measure the position of the bond density wave resonance $\omega_\mathrm{BDW}$. Firstly, as shown in Fig.~3(a) of the main text, via a linear fit of the density wave frequency response $\omega_\mathrm{DW}$, and, secondly, from the oscillation of the asymmetry of the momentum modes with negative and positive momentum with respect to the pump direction $F_\mathrm{+1,\pm1}\, - \,F_\mathrm{-1,\pm1}$. The third method is demonstrated in this paragraph by looking at the depletion of the cavity field in the parameter space spanned by the modulation frequency $\omega_\mathrm{dr}$ and modulation strength $f_\mathrm{0}$. On resonance the intracavity photon number $N_\mathrm{P}$ depletes fastest for a fixed $f_\mathrm{0}$. To quantify this effect, we divide the parameter space into $20\times16$ plaquettes. Then, we average $\sum N_\mathrm{P}$ in the interval from 2 to 3 ms after starting the modulation, and over multiple experimental runs. 

\begin{figure}[tbp]
\centering
\includegraphics[width=0.5\columnwidth]{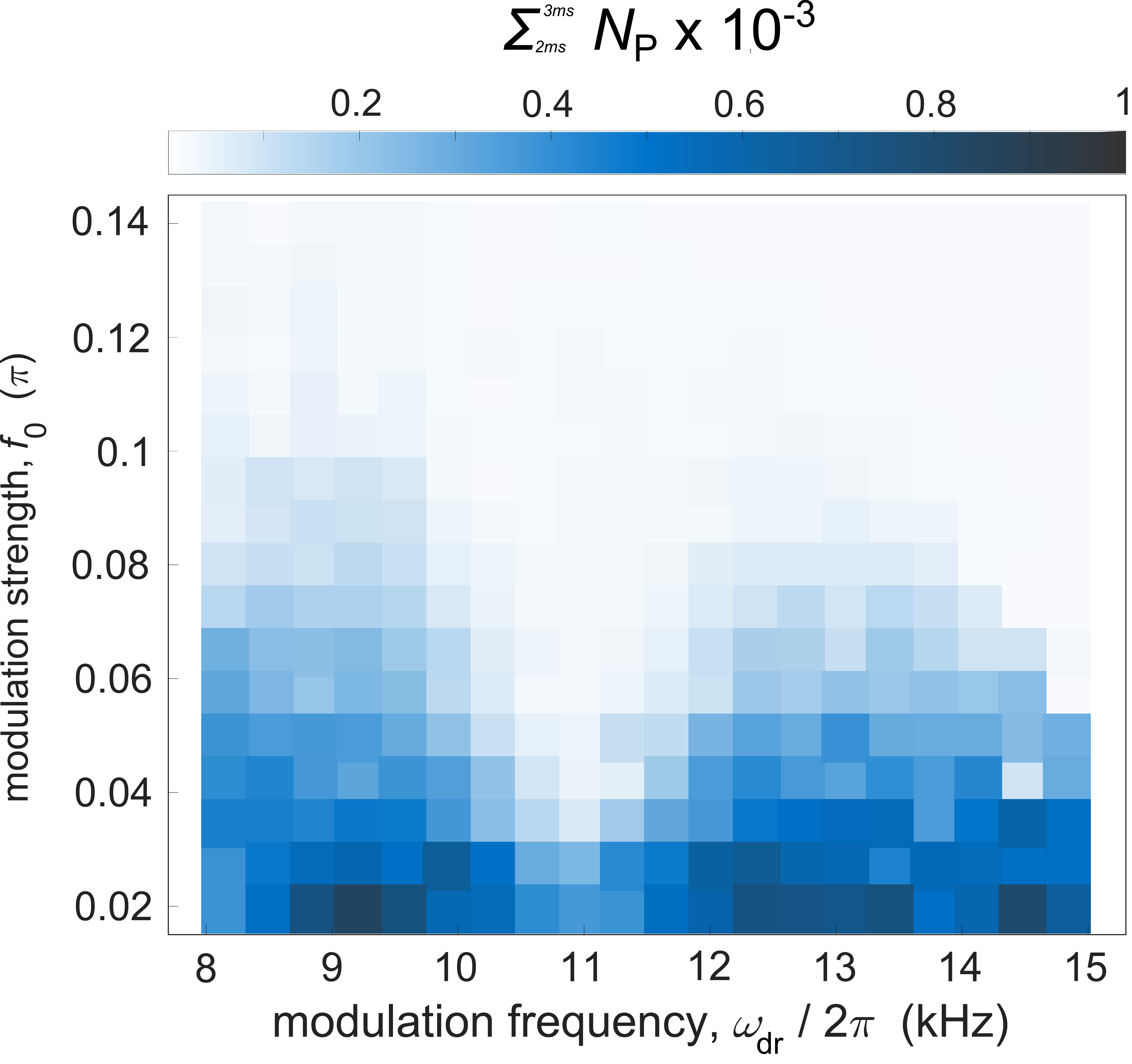}
\caption{{Sum of $N_\mathrm{P}$ in the interval from 2 to 3 ms.} We follow the protocol described in Fig.~3 of the main text for variable modulation frequencies $\omega_\mathrm{dr}$ and strengths $f_\mathrm{0}$. The color scale parametrizes the sum over the intracavity photon number $N_\mathrm{P}$ in the interval [2,3]~ms after reaching the final modulation strength $f_\mathrm{0}$.}
\label{sfig:2} 
\end{figure}

\section{Relation between the momentum imbalance and the density wave/bond density wave order parameters}
The density wave and bond density wave order parameters are defined as
\begin{align}
\Theta_\mathrm{DW}= \int \cos(ky)\cos(kz)|\psi(y,z)|^2 dx dy \\\
\Theta_\mathrm{BDW}=\int \sin(ky)\cos(kz)|\psi(y,z)|^2 dx dy~,
\end{align}
respectively. Using the spatial translation symmetry in the system we expand the atomic field in terms of plane waves as
\begin{equation}
\psi(y,z) = \sum_{n,m} \phi_{n,m} e^{i n y} e^{i m z}.
\end{equation}
With this the order parameter can be written in the momentum basis as
\begin{align}
\Theta_\mathrm{DW}&= \frac{1}{4} \sum_{n,m} \left( \phi^*_{n+1,m+1}+\phi^*_{n+1,m-1}+\phi^*_{n-1,m+1}+\phi^*_{n-1,m-1} \right) \phi_{n,m} \\
\Theta_\mathrm{BDW}&= \frac{1}{4i} \sum_{n,m} \left( \phi^*_{n+1,m+1}+\phi^*_{n+1,m-1}-\phi^*_{n-1,m+1}-\phi^*_{n-1,m-1} \right) \phi_{n,m}~.
\end{align}
In the following we will only retain the lowest five momentum modes $\{\phi_{0,0},\phi_{\pm,\pm}\}$. We approximate the order parameters as
\begin{align}
\Theta_\mathrm{DW}&= \frac{1}{4} \left( \phi^*_{+,+}+\phi^*_{+,-}+\phi^*_{-,+}+\phi^*_{-,-} \right) \phi_{0,0}+ \mathrm{h.c.} \\
\Theta_\mathrm{BDW}&= \frac{1}{4i} \left( \phi^*_{+,+}+\phi^*_{+,-}-\phi^*_{-,+}-\phi^*_{-,-} \right) \phi_{0,0}+ \mathrm{h.c.}~.
\end{align}
As the momentum modes along the $z$-direction will stay degenerate, we denote $ \phi_{+,\pm}=\phi_+$ and $ \phi_{-,\pm}=\phi_-$ and introduce the shorthand notation $\phi_{0,0} = \phi_0$. Then, the order parameter can be written as
\begin{align}
\Theta_\mathrm{DW}&= \frac{1}{2} \left( \phi^*_++\phi^*_- \right) \phi_0+ \mathrm{h.c.} \\
\Theta_\mathrm{BDW}&= \frac{1}{2i} \left( \phi^*_+-\phi^*_- \right) \phi_0+ \mathrm{h.c.}~.
\end{align}
The product of these two order parameters leads to
\begin{align}
\Theta_\mathrm{DW} \times \Theta_\mathrm{BDW}&=\frac{1}{4i}\left\{(\phi^*_0)^2\left((\phi_+)^2-\phi_-)^2 \right)-(\phi_0)^2\left((\phi^*_+)^2-\phi^*_-)^2 \right)+2|\phi_0|^2 \left( \phi^*_-\phi_+-\phi^*_+\phi_- \right) \right\} \\
&= \frac{|\phi_0|^2}{2} \left( |\phi_+|^2 \sin\left(2(\theta_+-\theta_0)\right)-|\phi_-|^2 \sin\left(2(\theta_--\theta_0)\right)+2|\phi_+||\phi_-|\sin\left(\theta_+-\theta_-\right) \right)~,
\end{align}
where we used in the last line $\psi_i= |\psi_i| \exp(i \,\theta_i)$. From our numerics, we find that this observable, as measured in the experiment, can be approximated by
\begin{equation}
\Theta_\mathrm{DW} \times \Theta_\mathrm{BDW} \approx \frac{|\phi_0|^2}{2} \left( |\phi_+|^2-|\phi_-|^2\right) \equiv \Delta F~.
\end{equation}

\begin{figure}[tbp]
\centering
\includegraphics[width=0.49\columnwidth]{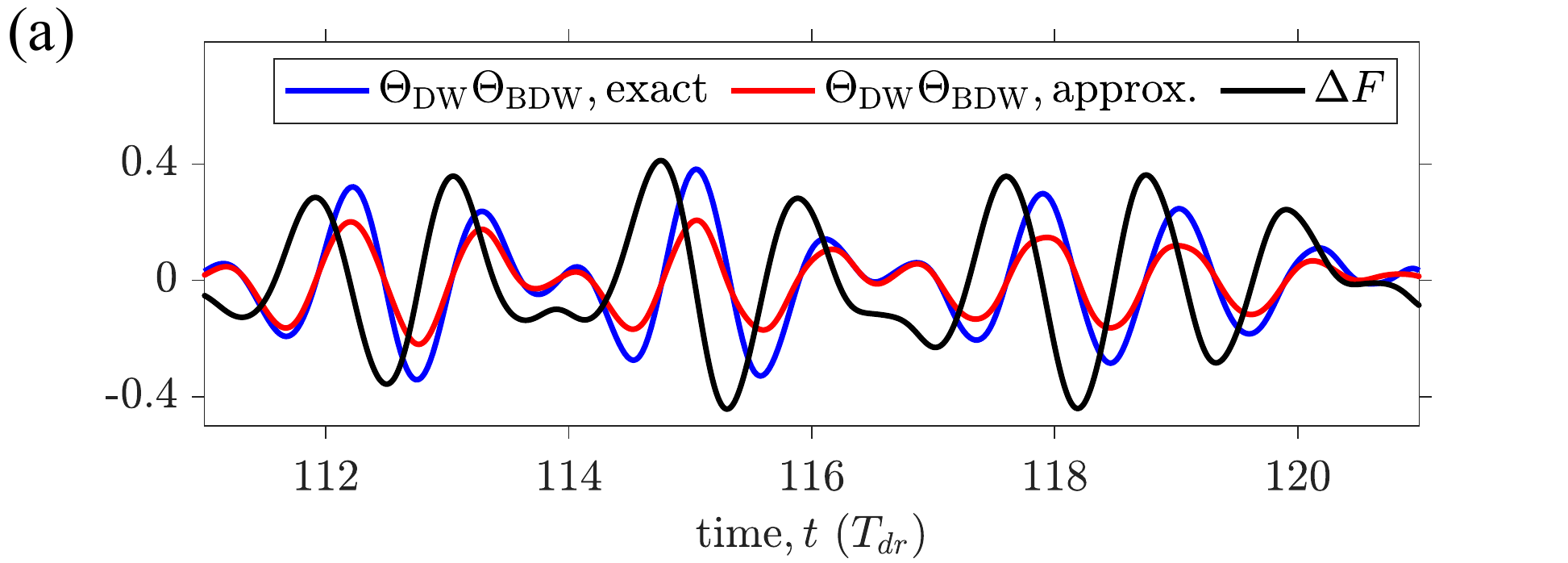}
\includegraphics[width=0.49\columnwidth]{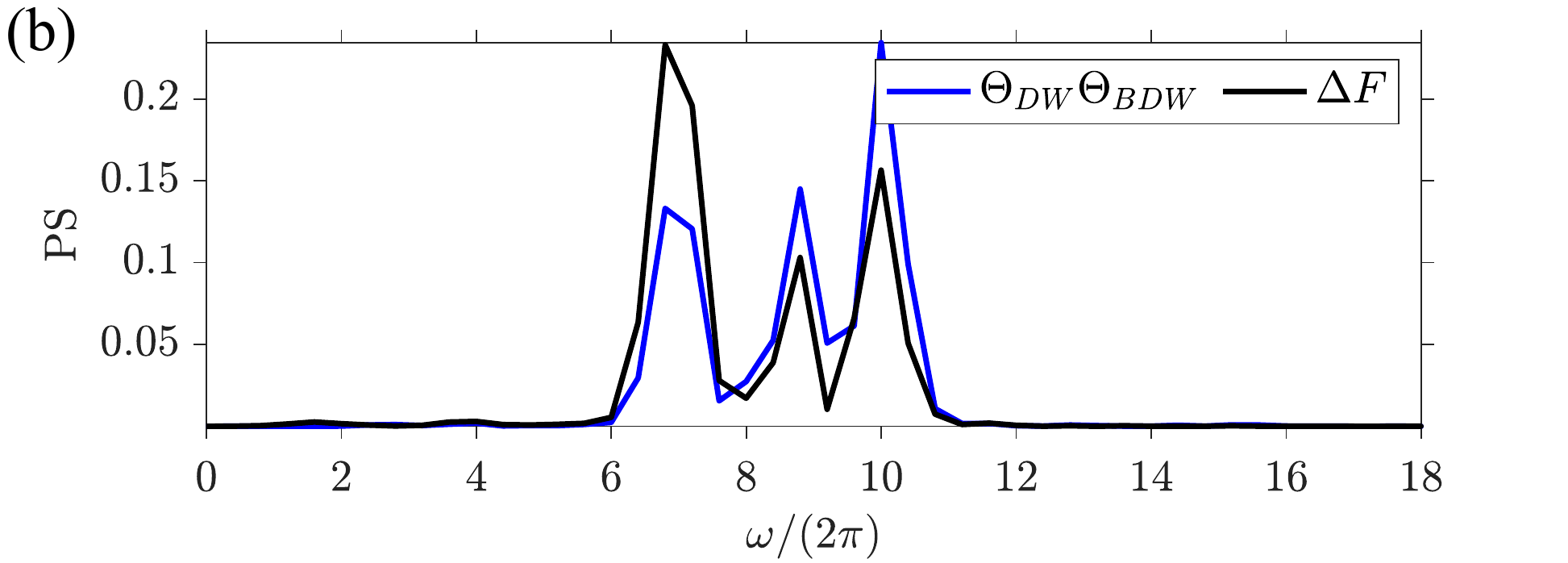}
\caption{Comparsion between the product of the two order parameters $\Theta_\mathrm{DW}\times\Theta_\mathrm{BDW}$ and the momentum imbalance $\Delta F$. The modulation strength is $f_0 = 0.1$. (cf. text)}
\label{sfig:3} 
\end{figure}

\begin{figure}[tbp]
\centering
\includegraphics[width=0.5\columnwidth]{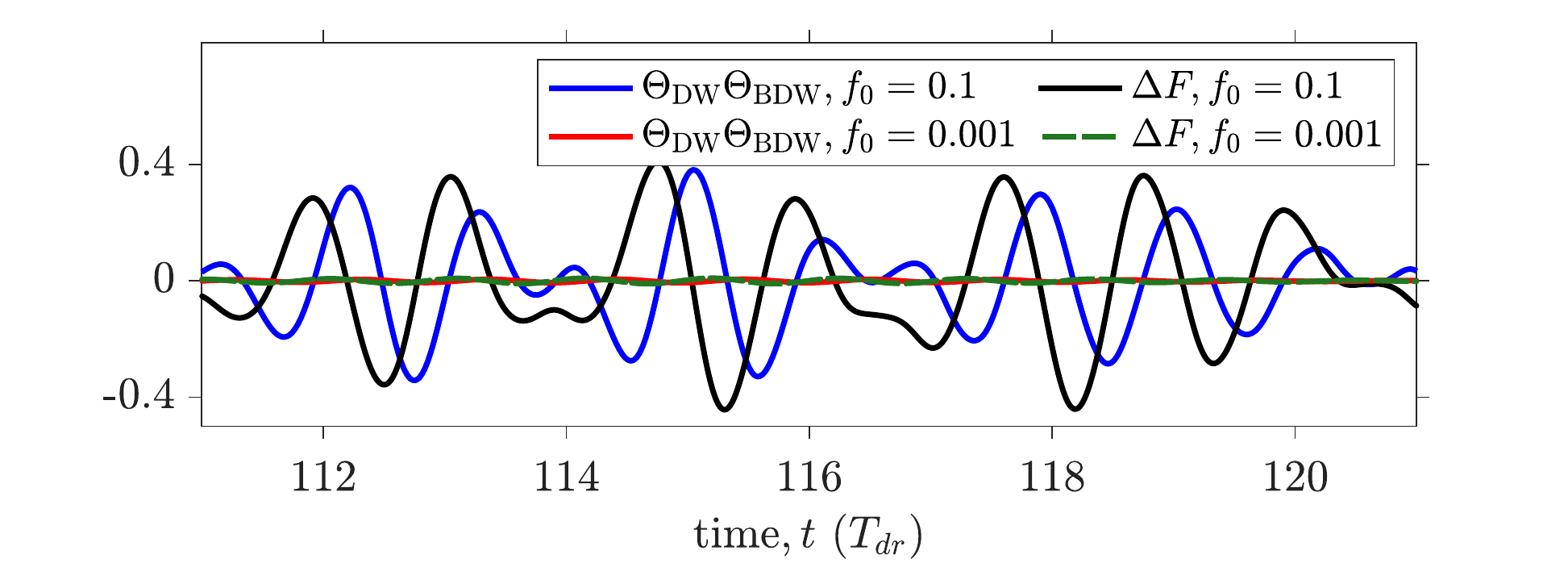}
\caption{Comparison of $\Theta_\mathrm{DW}\times\Theta_\mathrm{BDW}$ and $\Delta_F$ for driving strengths $f_\mathrm{0}=0.1$ and $f_\mathrm{0}=0.001$, respectively.}
\label{sfig:4} 
\end{figure}

Fig.~\ref{sfig:3} (a) shows a comparison between the product $\Theta_\mathrm{DW}\times\Theta_\mathrm{BDW}$ of the two order parameters and the momentum imbalance $\Delta F$. The blue trace shows the exact numerical result. The red trace shows an approximation, if only the five lowest momentum modes are accounted for. To further validate our findings, we compute in Fig.~\ref{sfig:3} (b) the power spectra of $\Theta_\mathrm{DW}\times\Theta_\mathrm{BDW}$ and $\Delta F$. Fig.~\ref{sfig:4} shows that for a small ($f_0 = 0.001$) driving strength $\Theta_\mathrm{DW}\times\Theta_\mathrm{BDW}$ and $\Delta F$ approach zero. Hence, the driven three-level Dicke model regime can only be realized for sufficiently strong driving.

\section{Lifetime of the time crystalline response}
\begin{figure}[tbp]
\centering
\includegraphics[width=0.5\columnwidth]{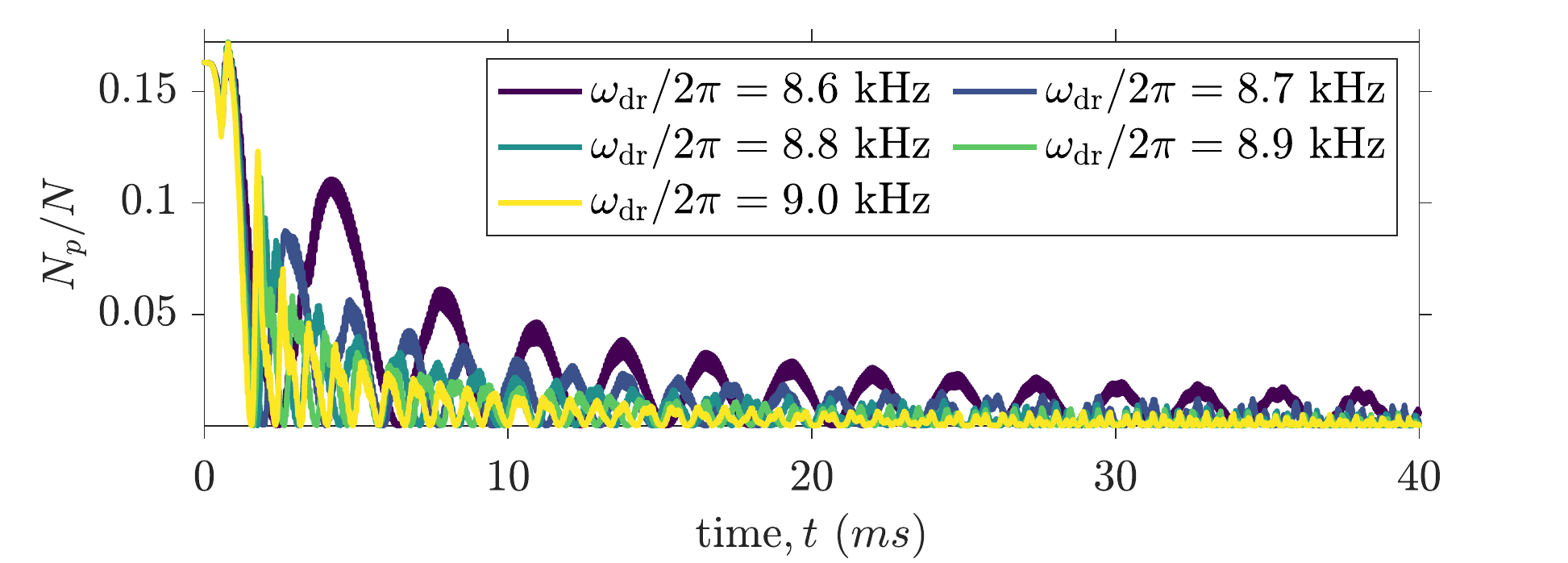}
\caption{{Decay of the time crystal response for different detunings} We choose a modulation strength of $f_\mathrm{0}=0.05$. The resonance frequency is located at $\approx 8.45~\mathrm{kHz}$. For larger detunings the time crystal melts more quickly.}
\label{sfig:5} 
\end{figure}

As was pointed out in \cite{Cosme2019}, the time crystalline response becomes unstable/pre-thermal as one scans further away from the resonance. Using mean-field theory without contact interactions, Fig.~\ref{sfig:5} shows the decay of the oscillations for different detunings of the driving frequency with respect to the resonance frequency. We note that this effect contributes to the decay observed in the experiment.

\section{Robustness against temporal perturbations.}
\begin{figure}[tbp]
\centering
\includegraphics[width=0.9\columnwidth]{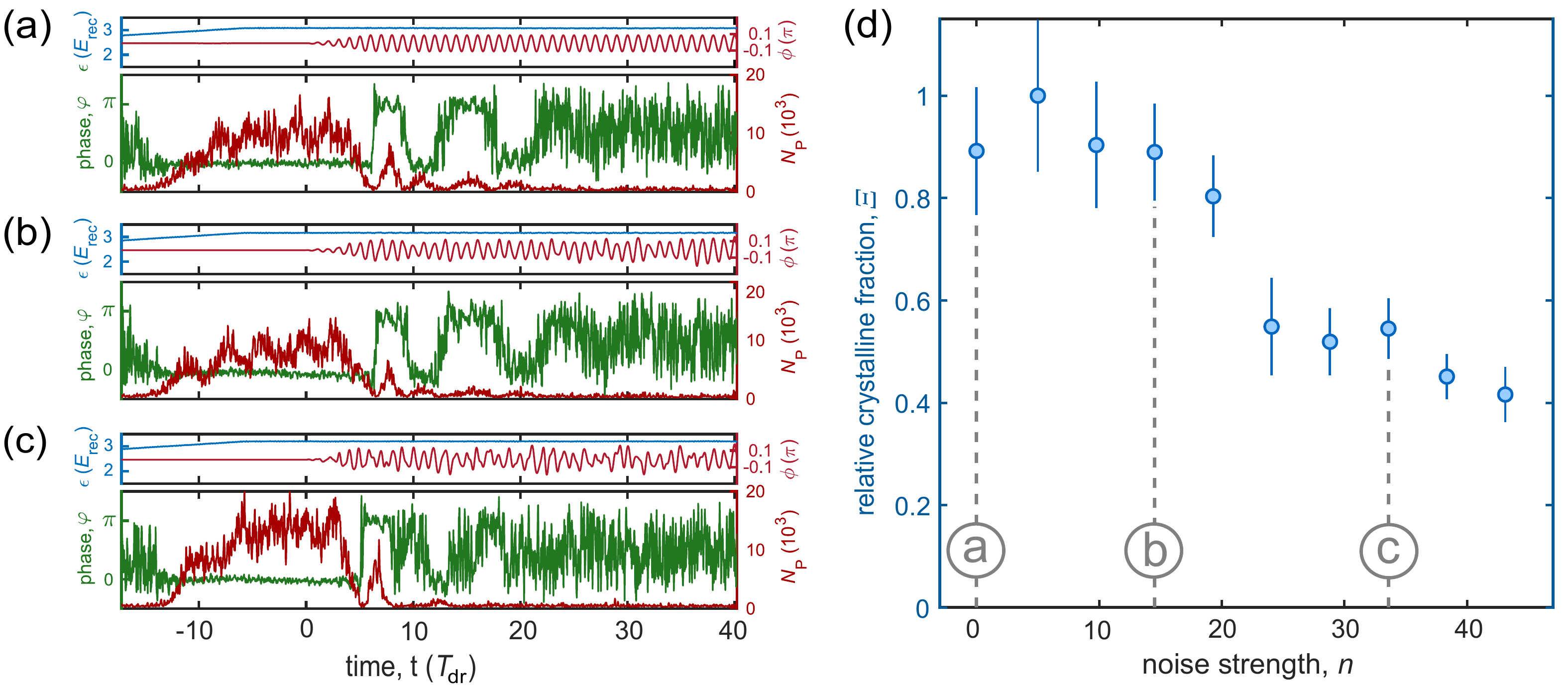}
\caption{{Robustness of the incommensurate time crystal.} (a)-(c) Single-shot experimental runs for the noise strengths marked in subplot (d) with the gray dashed lines. Top panels: single-shot protocols for the pump strength. Bottom panel: corresponding time evolution of the relative phase $\varphi$ (green trace) and intracavity photon number $N_\mathrm{P}$ (red trace). (d) Dependence of the relative crystalline fraction $\Xi$ on the noise strength averaged over 10 experimental runs with $f_0 = 0.1\,\pi$ and $\omega_\mathrm{dr} = 2\pi \times 11.5\,$kHz.}
\label{sfig:6} 
\end{figure}

The robustness against temporal perturbations is one of the main characteristics of time crystalline dynamics. We have tested the stability of the observed bond density wave phase against artificial white noise on the modulation signal with a bandwidth of $20\,$kHz. The applied noise strength is measured by $n \equiv \sum_{\omega=0}^{2\pi \times 20~\mathrm{kHz}} |\mathcal{E}_\mathrm{noisy}(\omega)|/\sum_{\omega=0}^{2\pi \times 20~\mathrm{kHz}} |\mathcal{E}_\mathrm{clean}(\omega)|$,  where $\mathcal{E}_\mathrm{noisy}$ ($\mathcal{E}_\mathrm{clean}$) is the Fourier spectrum of the pump in the presence (absence) of white noise. Figures~\ref{sfig:6}(d) show how the relative crystalline fraction $\Xi$ changes with increasing noise strength for fixed modulation parameters $f_\mathrm{0}=0.1$ and $\omega_\mathrm{dr}=11.5\,$kHz. Note that even for a strongly distorted pump signal, as in Fig.~\ref{sfig:6}(b) the system still switches multiple times between the two sublattices before the intracavity field disappears. 

\section{Single-shot realization of DBDW order for $\omega_\mathrm{dr}=12 \,$kHz}
\begin{figure}[tb]
\centering
\includegraphics[width=0.6\columnwidth]{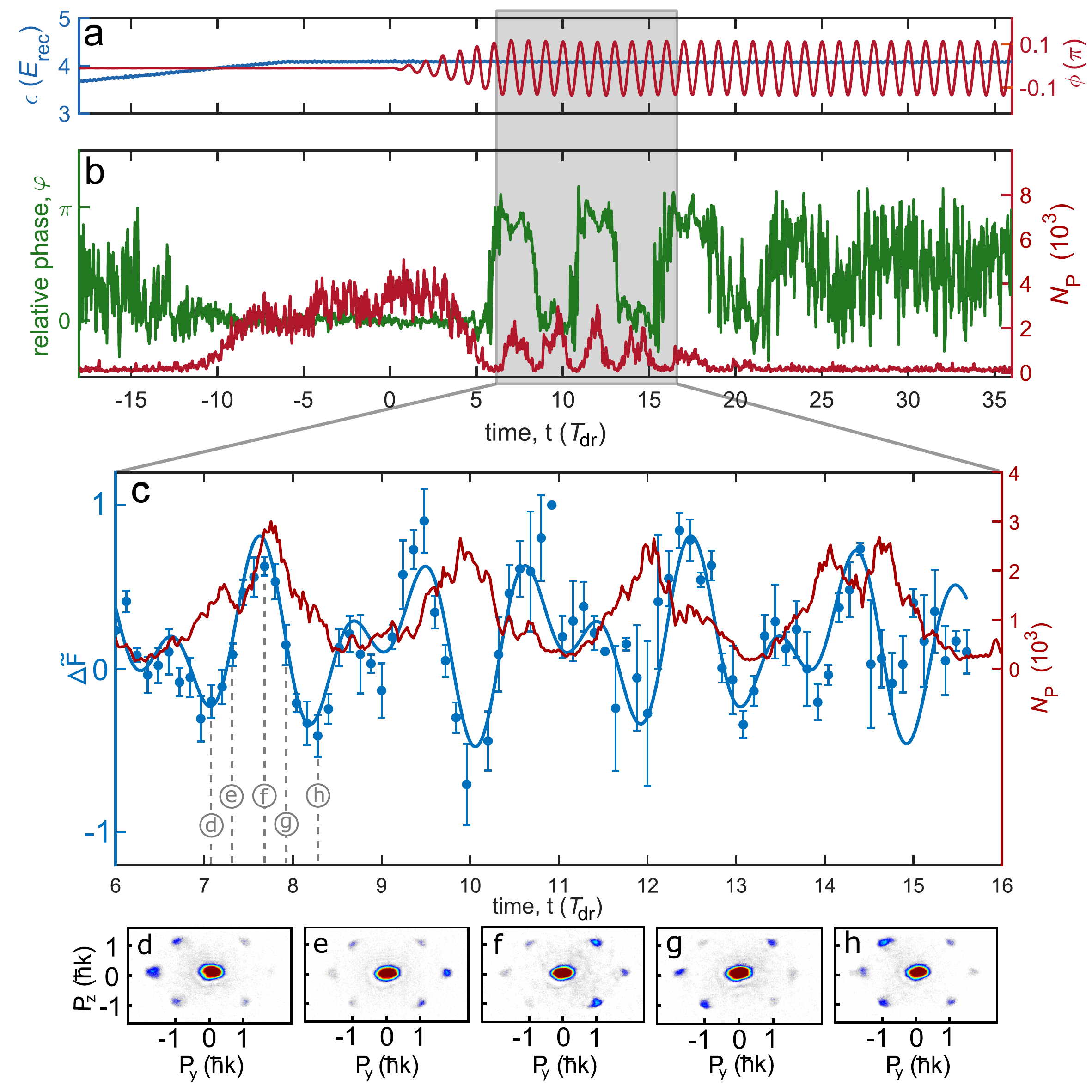}
\caption{{Single-shot realization of DBDW order.}  {(a)} Time sequence for the pump lattice depth (blue) and the phase $\phi$ of the pump field (red) with modulation strength $f_\mathrm{0}=$0.13$\,\pi$ and a modulation frequency $\omega_\mathrm{dr}=12.0\,$kHz.  {(b)} Phase difference $\varphi$ between the pump and intracavity field (green trace) and photon number $N_\mathrm{P}$ in the cavity (red trace). The dashed vertical lines mark the time interval during which the modulation strength is increased. The gray shaded area shows the time window for the zoom presented in (c).  {(c)} The red trace repeats the intracavity photon number $N_\mathrm{P}$ from (b). The blue data points plot the product $\Theta_{\mathrm{DW}} \times \Theta_{\mathrm{BDW}} $, approximately given by the difference between the number of atoms populating the momentum modes with positive and negative momentum with respect to the pump direction $\Delta \tilde{F} = (F_\mathrm{+1,\pm1}\,-\,F_\mathrm{-1,\pm1}) / (F_\mathrm{+1,\pm1}\,-\,F_\mathrm{-1,\pm1})_\mathrm{max}$ (see also Fig.~1(c) and Fig.~2(c) of the main text). Each data point is averaged over 5 realizations. The solid line shows a fit with a product of two harmonic oscillations. {(d)-(h)} Single-shot momentum distributions recorded at the times marked in (c).}
\label{sfig:7} 
\end{figure}

We measured the momentum mode asymmetry for a second parameter set and present the results in Fig.~\ref{sfig:7}. In Fig.~\ref{sfig:7}(c) we used $f(t)=\exp(-\tau t)\,A\sin(\omega_\mathrm{BDW}t+\alpha) \,\sin(\omega_\mathrm{DW}t)$ as a fit function. Here, $\tau$ is the decay rate of $N_\mathrm{P}$ and $A$ is an overall amplitude parameter. The fast BDW oscillation frequency is independent of $\omega_\mathrm{dr}$ and we measure $\omega_\mathrm{BDW} = 2\pi \times 9.8 \pm0.1\,$kHz. We find a slow oscillation frequency of $\omega_\mathrm{DW}=2\pi \times 2.6\pm0.1\,$kHz (see also Fig.~1(c) and Fig.~2(c) of the main text) for a driving frequency of $\omega_\mathrm{dr}=12.0\,$kHz, which agrees well with the theoretical prediction of $\omega_\mathrm{DW}=\omega_\mathrm{dr}-\omega_\mathrm{BDW}=2\pi\times(12.0 - 9.8)\,\mbox{kHz}=2\pi\times2.2\,$kHz.

\bibliography{references_BDW}